\documentclass[12pt,fleqn,epsfig]{article}
\usepackage{epsfig}
\usepackage{fleqn,espcrc1}
\usepackage{wrapfig}
\def\square{\vcenter{\vbox{\hrule height.4pt
          \hbox{\vrule width.4pt height8pt
          \kern8pt\vrule width.4pt}\hrule height.4pt}}}

\title{Free Energy of a Hot Gluon Plasma and Hard-thermal-loop Resummation}
 
\author{Jens O. Andersen,\address{Physics Department, Ohio State University, Columbus OH 43210, USA}\thanks{Talk given at 15th International Conference on Particle and Nuclei (PANIC 99), Uppsala, Sweden, 10-16 June 1999.}}
\begin{document}
 
\maketitle

\begin{abstract}
In this talk I briefly discuss the thermodynamics of the quark-gluon plasma
The calculation
of the free energy 
of a hot gluon plasma  to leading order in
hard-thermal-loop perturbation theory~\cite{EJM1} is outlined.  
The HTL free energy is compared with the weak-coupling expansion
and lattice results.
%
\end{abstract}
 \section{Introduction}
\begin{wrapfigure}{l}[0pt]{9cm}
\vspace{-0.8cm}
\epsfysize=7.3cm
\epsffile{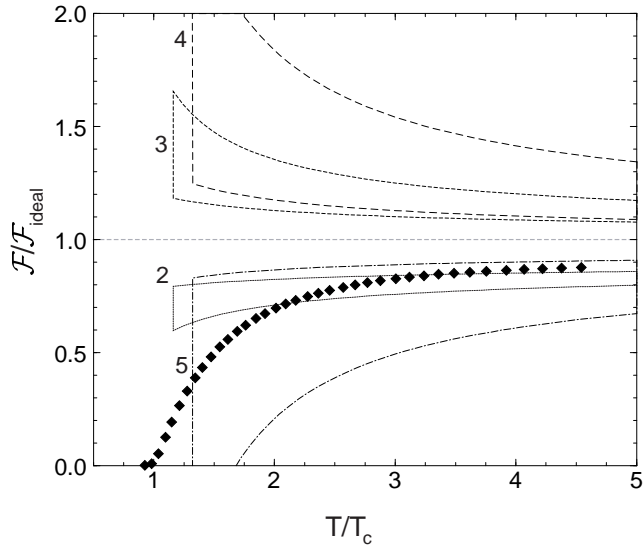}
\vspace*{-0.4cm}
\caption[a]{\small
The free energy for pure-glue QCD as a function of $T/T_c$.
See main text for details.}
\label{weakfig}
\vspace*{-0.7cm}
\end{wrapfigure}

Relativistic heavy-ion collisions will soon allow the experimental study of
hadronic matter at energy densities that will probably exceed 
that required to create a quark-gluon plasma.  
A quantitative understanding of the properties of a
quark-gluon plasma is essential in order to determine whether it has been
created. Because QCD is asymptotically free, the running coupling constant
$\alpha_s$ becomes small at sufficiently high temperatures. It therefore
seems plausible that one can understand the quark-gluon plasma using 
perturbative methods.
The free energy of QCD has been calculated to fifth order in the weak-coupling
expansion~\cite{Kastening-Zhai,Kastening-Zhai2,Braaten-Nieto}. For a pure-glue
plasma the result is
\begin{eqnarray}\nonumber
{\cal F}_{\rm QCD} &=& {\cal F}_{\rm ideal} 
\Bigg[ 1 - {15 \over 4} {\alpha_s \over \pi} 
+\ 30 \left ( {\alpha_s \over \pi} \right )^{3/2} 
+ {135 \over 2} \left( \log {\alpha_s \over \pi} 
-{11\over36}\log{\mu_4\over2\pi T}+ 3.51 \right) 
	\left( {\alpha_s \over \pi} \right)^2 
\nonumber
\\
&&
+{495\over2}\left(\log{\mu_4\over2\pi T}-3.23\right) \left ( {\alpha_s \over \pi} \right )^{5/2}
+ {\cal O} (\alpha_s^3\log \alpha_s ) \Bigg],
\label{F1-QCD}
\end{eqnarray}
where ${\cal F}_{\rm ideal}=-(8\pi^2/45)T^4$ is the free energy of an 
ideal gas of massless gluons and $\alpha_s=\alpha_s(\mu_4)$ is the running
coupling constant in the $\overline{\mbox{MS}}$ scheme.
In Fig.~\ref{weakfig}, the free energy is shown as a 
function of $T/T_c$, 
 where $T_c$ is the critical temperature
for the deconfinement phase transition.
The weak-coupling expansions through
orders $\alpha_s$, $\alpha_s^{3/2}$, $\alpha_s^2$, and $\alpha_s^{5/2}$
are shown as bands that correspond to varying the renormalization scale
$\mu_4$
by a factor of two from the central value $\mu_4=2\pi T$. 
The successive approximations to the free energy alternate in sign
and the perturbative expansion does not converge for temperatures that 
are relevant
for heavy-ion collisions. The $\alpha_s^{3/2}$ contribution is smaller
than the $\alpha_s$ contribution only if $T$ is larger than
$10^5$ GeV, while the temperatures 
at RHIC are expected not to exceed $T=0.5$ GeV.
The poor convergence of the perturbative series is somewhat surprising
since very accurate lattice results~\cite{lattice-0} indicate that 
the quark-gluon plasma can be approximated quite well by an ideal
gas unless the temperature is very close to $T_c$. The lattice results
for pure-glue QCD by Boyd et al.~\cite{lattice-0} are also
shown in Fig.~\ref{weakfig}
as diamonds. The free
 energy approaches that of an ideal gas from below
as $T/T_c$ increases.
\normalsize
\normalsize
There are several ways of reorganizing the perturbative series to improve
its convergence. One of the most successful approaches 
for scalar field theories is 
``screened perturbation theory''~\cite{karsch} which is an expansion
around an ideal gas of massive quasiparticles.
The convergence for the free energy is very good even for large values of 
the coupling constant.

Evidence that such a reorganization might be useful in QCD is provided by the
success of quasi-particle models. The analyses in~\cite{levai2} indicate
that lattice results for thermodynamic functions can be fit very well by 
a gas of massive quarks and gluons with temperature dependent masses
of order $T$. The quasi-particle models, however, suffer from several
theoretical inconsistencies because introducing thermal masses by hand
violates gauge invariance.

There is a way of incorporating plasma effects, such as quasi-particle masses,
screening of interactions, and Landau damping, into perturbative calculations,
while maintaing gauge invariance, and that is by using hard-thermal-loop (HTL)
perturbation theory.
HTL perturbation theory was originally developed to sum
up all higher loop corrections that are leading order in $g$ for amplitudes
having soft external lines with momenta of order $gT$~\cite{bp}. 
If it is applied
to amplitudes with hard external lines with momenta of order $T$, it 
selectively resums corrections that are higher order in $g$.
This resummation 
is a generalization of screened
perturbation theory that respects gauge invariance.
It corresponds to expanding around an
ideal gas of massive quark and gluon quasiparticles. 

A significant advantage of HTL perturbation theory over the approach
using lattice gauge theory
calculations is that it can be readily applied to the real-time
processes that are the most promising signatures of a quark-gluon plasma.
Due to the Euclidean formulation, lattice methods are  
restricted to calculating static quantities such as the free 
energy~\cite{lattice-0}
and the Debye mass~\cite{debye}.
\section{HTL Free Energy} 
In the imaginary time formalism, 
the one-loop HTL free energy for an $SU(3)$ gauge theory is~\cite{EJM1}
\begin{eqnarray}
\label{freedef}
{\cal F}_{\rm HTL}=8\left[(d-1){\cal F}_T+{\cal F}_L+\Delta{\cal F}\right],
\end{eqnarray} 
where
\begin{eqnarray} 
{\cal F}_T &=&  {1 \over 2}\sum_{n=-\infty}^{\infty}\int{d^{d}k\over(2\pi)^{d}}
\log [k^2 +\omega_n^2+ \Pi_T(\omega_n,k)], 
\label{FT-def}\\
{\cal F}_L  &=&  {1 \over 2}\sum_{n=-\infty}^{\infty}
\int{d^{d}k\over(2\pi)^{d}}
\log [k^2 - \Pi_L(\omega_n,k)],
\label{FL-def}
\end{eqnarray} 
and $\Delta \cal F$ is a counterterm.
Dimensional regularization in $d=3-2\epsilon$ dimensions
is used throughout.
The HTL self-energy functions $\Pi_T$ and $\Pi_L$ are
\begin{eqnarray}
\Pi_T(\omega_n,k) & = & - {3 \over 2} m_g^2 {\omega_n^2 \over k^2} 
\left[1 + {\omega^2_n + k^2 \over 2i\omega_nk} 
	\log{i\omega_n+k \over i\omega_n - k} \right],
\label{Pi-T}\\
\Pi_L (\omega_n,k) &=& 3m_g^2 \left[{i \omega_n \over 2k} 
	\log {i\omega_n +k \over i \omega_n-k} - 1 \right],
\label{Pi-L}
\end{eqnarray}
and $m_g$ is the gluon mass parameter.
The sum over the Matsubara frequencies $\omega_n=2\pi nT$ can be rewritten as a contour integral
around a contour $C$ that encloses the points $\omega=i\omega_n$. 
The integrand has branch cuts that start at $\pm\omega_T(k)$ and
$\pm\omega_L(k)$, where $\omega_T(k)$ and $\omega_L(k)$
are the dispersion relations for transverse and longitudinal 
gluon quasiparticles, respectively. The integrand also
has a branch cut running from $\omega=-k$ to $\omega=k$ due to the functions
$\Pi_T$ and $\Pi_L$.
The contour
can be deformed to wrap around the quasiparticle and Landau-damping
branch cuts. Some of the temperature-independent integrals over $\omega$ 
can be
calculated analytically, while others must be evaluated numerically. 
With dimensional regularization, the logarithmic ultraviolet 
divergences show up as poles in $\epsilon$.
Using the modified minimal subtraction ($\overline{\mbox{MS}}$)
renormalization prescription
with the counterterm $\Delta{\cal F}=9m^4_g/64\pi^2\epsilon$, we obtain
\begin{eqnarray} \nonumber
{\cal F}_{\rm HTL}&=&{4T\over\pi^2}\int_0^{\infty}
k^2dk\left[2\log(1-e^{-\beta\omega_T})
+\log{1-e^{-\beta\omega_L}\over1-e^{-\beta k}}\right] \\ 
&&
+{4\over\pi^3}\int_0^{\infty}{d\omega\over e^{\beta\omega}-1}
\int_{\omega}^{\infty}k^2dk\left[\phi_L-2\phi_T\right]
\label{total}
+{1\over2}m_g^2T^2+{9m_g^4\over8\pi^2}
\left[\log{m_g\over\mu_3}-0.33\right].
\end{eqnarray} 

\begin{wrapfigure}[]{l}[0pt]{9cm}
\vspace{-0.7cm}
\epsfysize=7.3cm
\epsffile{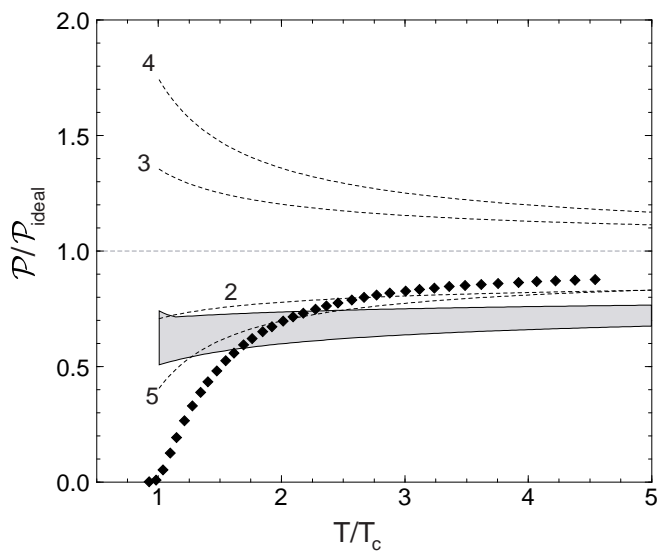}
\vspace*{-0.3cm}
\caption[a]{\small
The HTL pressure for pure-glue QCD as a function of
$T/T_c$. 
See main text for details.}
\label{pres}
\vspace*{-0.7cm}
\end{wrapfigure}
$\mu_3$ is the renormalization scale associated with dimensional
regularization, and the angles $\phi_T$ and $\phi_L$ are 
complicated functions of 
$\omega$ and $k$~\cite{EJM1}.

The leading-order HTL result for the pressure is shown in 
Fig.~\ref{pres} as the shaded band that corresponds to varying the
renormalization scales $\mu_3$ and $\mu_4$ 
by a factor of two around their central values
$\mu_3=0.717m_g$ and $\mu_4=2\pi T$.
This value of $\mu_3$ is chosen in order to minimize the 
pathological behavior of ${\cal F}_{HTL}$ at low temperatures~\cite{EJM1}.
We also show as dashed curves the weak-coupling expansions through order
$\alpha_s$, $\alpha_s^{3/2}$, $\alpha_s^2$, and $\alpha_s^{5/2}$ labelled
2, 3, 4, and 5.
We have used a parametrization of the running coupling constant 
$\alpha_s(\mu_4)$ that includes
the effects of two-loop running
With the above choices of the renormalization scales, 
our leading-order result for the HTL free energy lies 
below the lattice results of Boyd et al.~\cite{lattice-0} 
(shown as diamonds) for $T>2T_c$.
However, the deviation from lattice QCD results
has the correct sign and roughly the correct magnitude to be
accounted for by next-to-leading order corrections in HTL perturbation 
theory~\cite{EJM1}.
This can be seen from the
high-temperature expansion of the free energy, which is an expansion 
in powers of $m_g/T$.
To order $m^4_g/T^4$ one obtains
\begin{eqnarray}
{\cal F}_{\rm HTL} &=& {\cal F}_{\rm ideal} 
\Bigg[ 1 - {45 \over 4} \left( {3 m_g^2 \over 4 \pi^2 T^2} \right) 
+ 30  \left( {3 m_g^2 \over 4 \pi^2 T^2} \right)^{3/2}
\nonumber
\\
&& 
+  {45 \over 8} \left( \log {\mu_3^2 \over 4 \pi^2 T^2} - 1.232 \right)
	\left( {3 m_g^2 \over 4 \pi^2 T^2} \right)^2 
+ {\cal O} (m_g^6 / T^6) \Bigg].
\label{FHTL-high}
\end{eqnarray} 
Comparing (\ref{FHTL-high}) with the weak-coupling expansion~(\ref{F1-QCD})
and identifying $m_g^2$ with its weak-coupling limit 
${4\pi\over3}\alpha_sT^2$, 
we conclude
that HTL perturbation theory overincludes the $\alpha_s$ contribution
by a factor of three. The $\alpha_s^{3/2}$ contribution 
which is associated with Debye screening is included correctly. 
At next-to-leading order, HTL perturbation theory agrees with the weak-coupling
expansion through order $\alpha_s^{3/2}$. 
Thus the next-to-leading order contribution to
${\cal F}_{}/{\cal F}_{\rm ideal}$ in HTL perturbation theory will be positive
at large $T$ since it must approach $+{15\over2}\alpha_s/\pi$.
\\

The talk is based upon work with Eric Braaten and 
Michael Strickland~\cite{EJM1}. 
The author would like to thank the organizers of PANIC 99 for a stimulating
meeting. 
This work was supported in part by a Faculty Development Grant 
from the Physics Department of the Ohio State University
and by a fellowship from the Norwegian Research Council 
(project 124282/410).

\end{document}